\documentclass[aps,prb,twocolumn,superscriptaddress,showpacs]{revtex4}

\usepackage{hyperref,latexsym}
\usepackage{graphicx,color,pstricks}
\usepackage{epsfig,epsf,bm}

\begin{document}
\title{The Stability of the Low-dimensional Mixtures of Dilute Quantum
Gases}
\author{Yu-Li Lee}
\affiliation{Department of Physics, National Changhua University of
Education, Changhua, Taiwan, R.O.C.}
\author{Yu-Wen Lee}
\affiliation{Department of Physics, Tunghai University, Taichung, Taiwan,
R.O.C.}

\begin{abstract}
 We study the stability of the dilute Bose-Fermi and
 Bose-Bose mixtures with repulsive interactions in one and two
 dimensions in terms of the renormalization group. For the
 Bose-Fermi mixture, we show that the uniform mixture is stable
 against de-mixing in the dilute limit. For the Bose-Bose mixture,
 we give the stability conditions in the dilute limit. As a
 byproduct, we also calculate the critical temperature for the
 superfluid phase of the two-dimensional Bose-Fermi mixture in the
 extremely dilute limit.
\end{abstract}



\maketitle


\section{Introduction}

Ever since the achievement of Bose-Einstein condensation in the
dilute gases of ultracold atoms, the study of degenerate quantum
gases has become a subject attracting a lot of experimental and
theoretical vigor\cite{PS}. Due to the experimental facilities
which allow control of various parameters characterizing the
system, it provides an ideal laboratory for the studies of quantum
many-body physics. Among them, one line of researches is to study
the mixture of quantum gases. These include the mixture of bosonic
and fermionic atoms (the Bose-Fermi mixture)\cite{BF}, and the
mixture of two different hyperfine states of the same bosonic
atoms\cite{BB1} or two different kinds of bosonic atoms\cite{BB2}
(the Bose-Bose mixture). The study of the degenerate mixtures of
quantum gases at lower dimensions is particularly interest and a
lot of theoretical investigations are devoted to
it\cite{Das,CH,MWHLD,bethe,RB,RI,MJP,2dbf,AHDL,MPD}. This is because
the competition between the inter-species and intra-species
interactions and the strong quantum fluctuations in low dimensions
may introduce many interesting phenomena such as phase transition,
new quantum states, and quantum phase transitions.

The present work is to study the stability of the low-dimensional
Bose-Fermi (BF) and Bose-Bose (BB) mixtures with repulsive
interactions in the dilute limit. For binary mixtures with mutual
repulsions, the system may experience a de-mixing transition by
varying the strength of the interspecies interactions. The
de-mixing transition or the stability of the one-dimensional
($1$D) uniform mixture has been analyzed by using either the
mean-field theory\cite{Das} or the bosonization\cite{CH,MWHLD}.
The mean-field theory has the well-known problem that it can not
treat properly the strong fluctuations in low dimensions. Though
there is nothing wrong with the bosonization approach in one
dimension, a naive extrapolation of the results obtained by the
bosonization to the strong-coupling regime may sometimes lead to
incorrect conclusions. For example, the bosonization approach
predicts that the BF mixture will be phase separated in the strong
repulsion regime\cite{CH}. In the solvable limit, however, the
Bethe-ansatz result indicates that the uniform BF mixture is
stable against de-mixing irrespective of the strength of
repulsions\cite{bethe}. Therefore, it is desirable to study this
issue from a different approach. For the $1$D single-component
dilute Bose gas, it has been shown that the low-energy physics in
the strong coupling regime can be well-captured by the
renormalization-group (RG) method starting from a zero-density
quantum critical point (QCP)\cite{SSS}. Here we extend this idea
to the low-dimensional mixtures with repulsions. Our results should
provide a point of view complementary to the bosonization approach
in the strong-coupling limit.

\begin{figure}
 \begin{center}
 \includegraphics[width=0.9\columnwidth]{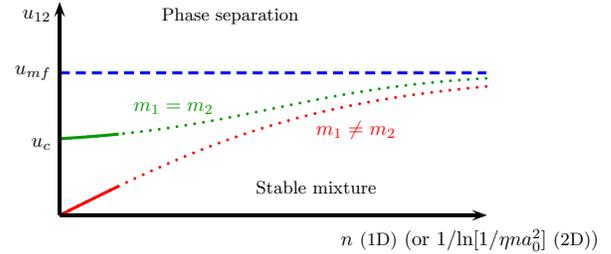}
 \end{center}
 \caption{(color online) The schematic phase diagram of the $1$D ($2$D)
 BB mixture with repulsions. We have taken $n_1=n/2=n_2$. The dashed
 line is the mean-field value. In $1$D, the critical strength
 ${\mathcal U}_c^{(1)}$ at low density for $m_1\neq m_2$ is a linear
 function of $n$. In $2$D, the critical strength ${\mathcal U}_c^{(2)}$
 at extremely low density for $m_1\neq m_2$ is a linear function of
 $1/\ln{[1/(\eta na_0^2)]}$, where $\eta =\frac{m_2/m_1}{2\pi (1+m_2/m_1)^2}$.
 For $m_1=m_2$, both ${\mathcal U}_c^{(1)}$ and ${\mathcal U}_c^{(2)}$
 take a constant value $u_c$ as $n\rightarrow 0$. Moreover,
 ${\mathcal U}_c^{(1)}$ and ${\mathcal U}_c^{(2)}$ will approach the
 mean-field value $u_{mf}=\sqrt{u_{11}u_{22}}$ at large density. In
 between, we assume that they are smooth functions. In the SU($2$)
 symmetrical limit, $u_c=u_{mf}$.}
 \label{1dbbp}
\end{figure}

In one dimension, there is an interacting zero-density fixed point
which controls the low-energy physics of the $1$D uniform mixture
in the strongly repulsive limit. Similar to the single-component
dilute Bose gas, this nontrivial fixed point can be captured by the
$\epsilon$ expansion, where $\epsilon =2-d$. For the two-dimensional
($2$D) mixtures, all couplings are marginally irrelevant in the sense
of RG. Thus, we expect that the stability conditions given by the
mean-field theory are correct to the leading order. Our main results
are as follows: (i) We find that the $1$D uniform BF mixture in the
dilute limit, i.e. the strong-coupling regime, is always stable against
de-mixing for any finite strength of repulsive interactions. This is
consistent with Bethe-ansatz result in the solvable limit\cite{bethe}.
And our analysis extends this result to the region beyond the solvable
limit. (ii) We derive the stability condition of the $1$D uniform BB
mixture in the strong-coupling regime (Fig. \ref{1dbbp}), which is
different from the bosonization result. (iii) In the extremely dilute
limit, we find that the interactions for the $2$D mixtures are strongly
renormalized so that the stability conditions obtained by the mean-field
theory are changed considerably. (iv) We also calculate the critical
temperature for the $2$D BF mixture in the extremely dilute limit [Eq.
(\ref{2dbfmtc31})] to investigate one aspect of the effects arising from
the inclusion of fermions. Although the stability of the mixtures with
attractions is also an interesting issue\cite{RI}, it is beyond the scope
of the present work. This is because the fixed-point structure for
attractions is different from the one for repulsions, i.e. the former
exhibits a runaway RG flow, and thus our approach cannot give definite
results. Finally, we notice that a similar RG analysis of the stability
of the dilute BB mixture in low dimensions was also carried out in a
recent paper\cite{K}. The results are different from ours in certain
aspects. We will address the origin of this discrepancy at the end of
the paper.

The rest of the paper is devoted to the detailed analysis that
leads to the above conclusions. Sec. \ref{bfm} and \ref{2bm} are
about the BF and BB mixtures, respectively. We compare our results
with the previous work in the last section. An appendix about the
mean-field treatment of the BF mixture in $d$ dimensions is
provided for reference.

\section{The Bose-Fermi mixture}
\label{bfm}

To study the mixture of bosons and spinless fermions in the dilute
limit,  we start with the action
\begin{eqnarray}
 S &=& \int^{\beta}_0 \! \! d\tau \! \! \int \! \! d^dx\psi^{\dagger} \! \!
 \left(\partial_{\tau}-\frac{\nabla^2}{2m_f}-\mu_f\right) \! \psi
 \nonumber \\
 & & +\int^{\beta}_0 \! \! d\tau \! \! \int \! \! d^dx\phi^{\dagger} \! \!
 \left(\partial_{\tau}-\frac{\nabla^2}{2m_b}-\mu_b\right) \! \phi \nonumber
 \\
 & & +\int^{\beta}_0 \! \! d\tau \! \! \int \! \! d^dx \! \left(\frac{g_{bb}}
 {2}|\phi|^4+g_{bf}|\phi|^2|\psi|^2\right) , ~~\label{bfms1}
\end{eqnarray}
where $\beta =1/T$. (We will take $\hbar =1$ and $k_B=1$.) Here
$\phi$ and $\psi$ are the fields describing the bosonic and
fermionic atoms, respectively. $m_f$, $\mu_f$ and $m_b$, $\mu_b$
are the masses and chemical potentials for the bosons and
fermions, respectively. The couplings $g_{bf}$ and $g_{bb}$ are
related to the scattering lengths $a_{bf}$ and $a_{bb}$ through
the relations $g_{bf}=2\omega_{\perp}a_{bf}$ and
$g_{bb}=2\omega_{\perp}a_{bb}$ for the $1$D trap,\cite{1d} and
$g_{bf}=\sqrt{4\pi(m_b+m_f)\omega_{\perp}/m_bm_f}a_{bf}$ and
$g_{bb}=\sqrt{8\pi\omega_{\perp}/m_b}a_{bb}$ for a $2$D
trap,\cite{2d} where $\omega_{\perp}$ is the transverse confining
frequency. We shall take $g_{bb},g_{bf}>0$ to avoid the possible
formation of bound states. We also assume that
\begin{equation}
 \mu_b ,\mu_f\ll\frac{1}{m_{\alpha}a_0^2} \ , \label{dilute1}
\end{equation}
where $\alpha =b,f$ and $a_0$ is the average range of interactions
between atoms, which can be regarded as the short-distance cutoff
of the action $S$. We will see later that this condition amounts
to the dilute limit.

The action $S$ [Eq. (\ref{bfms1})] has a QCP at $\mu =0$, $T=0$,
and $g_{bb}=0=g_{bf}$ (the Gaussian fixed point). In the dilute
limit, this Gaussian fixed point may be an appropriate departure
point to study the low-energy physics of this system. To proceed,
we shall employ the RG method. The strategy is as the following.
We first analyze the fixed-point structure of the theory. Then we
integrate the RG equations to the scale where the correlation
length becomes of $O(1)$ in the unit of the cutoff of the resulting
renormalized Hamiltonain. At this scale the quantum fluctuations
have been taken into account (within the $\epsilon$ expansion), and
thus we may apply the mean-field theory to the renormalized Hamiltonian
to study the stability of the mean-field solution corresponding to
the uniform mixture. In terms of the solutions of RG equations, we may
express the stability conditions in terms of the physical quantities.

\subsection{Renormalization-group theory}

To proceed, we first make a change of variables:
\begin{eqnarray*}
 \bm{x}=a_0\tilde{\bm{x}} \ , & & \tau =ma_0^2\tilde{\tau} \ ,
 \\
 \psi =a_0^{-d/2}\tilde{\psi} \ , & & \phi =a_0^{-d/2}\tilde{\phi} \ ,
\end{eqnarray*}
such that $\tilde{\bm{x}}$, $\tilde{\tau}$, $\tilde{\psi}$,
$\tilde{\phi}$ all become dimensionless, where $m$ is a quantity carrying
the dimension of mass. Then, the action $S$ [Eq. (\ref{bfms1})] at $T=0$
can be written as
\begin{eqnarray*}
 S &=& \int \! \! d\tilde{\tau} \! \! \int \! \! d^d\tilde{x}
 \tilde{\psi}^{\dagger} \! \! \left(\partial_{\tilde{\tau}}-\frac{\nabla^2}
 {2\tilde{m}_f}-r_f\right) \! \tilde{\psi} \\
 & & +\int \! \! d\tilde{\tau} \! \!
 \int \! \! d^d\tilde{x}\tilde{\phi}^{\dagger} \! \! \left(
 \partial_{\tilde{\tau}}-\frac{\nabla^2}{2\tilde{m}_b}-r_b\right) \!
 \tilde{\phi} \\
 & & +\int \! \! d\tilde{\tau} \! \! \int \! \! d^d\tilde{x} \! \left[
 \frac{\lambda_{bb}}{2K_d\tilde{m}_b}|\tilde{\phi}|^4+\frac{\lambda_{bf}}
 {K_d\tilde{m}_{bf}}|\tilde{\phi}|^2|\tilde{\psi}|^2\right] ,
\end{eqnarray*}
where $m_{bf}=2m_bm_f/(m_b+m_f)$, $\tilde{m}_b=m_b/m$, $\tilde{m}_f=m_f/m$,
$\tilde{m}_{bf}=m_{bf}/m$, and
\begin{eqnarray}
 r_f=m\mu_fa_0^2 \ , & & r_b=m\mu_ba_0^2 \ , \nonumber \\
 \lambda_{bb}=K_dm_bg_{bb}a_0^{\epsilon} \ , & & \lambda_{bf}=K_dm_{bf}
 g_{bf}a_0^{\epsilon} \ , \label{para1}
\end{eqnarray}
are all dimensionless parameters. In the above, we define
$K_d=2/[(4\pi)^{d/2}\Gamma (d/2)]$ ($\Gamma (x)$ is the Gamma
function) and $\epsilon =2-d$. We will see later that $m$ always
appear in the physical quantities in the guise of $ma_0^2$.
Therefore, for universal quantities which do not depend on $a_0$
explicitly, the dependence on $m$ is completely dropped out. The
determination of the short-distance energy scale $1/(ma_0^2)$
relies on the comparison with exact solutions or experimental
results.

To perform the RG transformations, we decompose the fields
$\tilde{\psi}$ and $\tilde{\phi}$ into
$\tilde{\psi}=\tilde{\psi}_<+\tilde{\psi}_>$ and
$\tilde{\phi}=\tilde{\phi}_<+\tilde{\phi}_>$, where
$\tilde{\psi}_>$ ($\tilde{\phi}_>$) consists of the Fourier
components of $\tilde{\psi}$ ($\tilde{\phi}$) with $\Lambda
e^{-l}<|\bm{k}|<\Lambda$, and $\tilde{\psi}_<$ ($\tilde{\phi}_<$)
consists of the Fourier components of $\tilde{\psi}$
($\tilde{\phi}$) with $|\bm{k}|<\Lambda e^{-l}$. Here $\Lambda$ is
the (dimensionless) UV cutoff for the momentum and $l>0$ is the
scaling parameter. By integrating out $\tilde{\psi}_>$ and
$\tilde{\phi}_>$ to the one-loop order and then rescaling the
momentum and frequency by $\bm{k}\rightarrow e^{-l}\bm{k}$ and
$\omega\rightarrow e^{-2l}\omega$, we obtain an effective action
for the slow modes. By ignoring the irrelevant operators in the
sense of RG, this effective action has an identical form to the
original action $S$ but with different values of the parameters
$r_{\alpha}(l)$, $\lambda_{bb}(l)$, and $\lambda_{bf}(l)$, which
can be regarded as the corresponding parameters at the momentum
scale $\Lambda e^{-l}$. By comparing the effective action of the
slow modes with the original action and considering an
infinitesimal value of $l$, we get the one-loop RG equations
\begin{equation}
 \frac{dr_{\alpha}}{dl}=2r_{\alpha} \ , \label{bfmrg21}
\end{equation}
and
\begin{eqnarray}
 \frac{d\lambda_{bb}}{dl} &=& \epsilon\lambda_{bb}-\frac{\lambda_{bb}^2}
 {1-2\tilde{m}_br_b} \ , \nonumber \\
 \frac{d\lambda_{bf}}{dl} &=& \epsilon\lambda_{bf}-\frac{\lambda_{bf}^2}
 {1-\tilde{m}_{bf}(r_b+r_f)} \ , \label{bfmrg22}
\end{eqnarray}
where the initial values $r_{\alpha}(0)$, $\lambda_{bb}(0)$, and
$\lambda_{bf}(0)$ are given by Eq. (\ref{para1}). Since we are
only interested in the dilute limit so that $r_{\alpha}(l)\ll 1$
before scaling stops, Eq. (\ref{bfmrg22}) can be approximated
as\cite{dilute}
\begin{equation}
 \frac{d\lambda}{dl}=\epsilon\lambda -\lambda^2 \ , \label{bfmrg23}
\end{equation}
where $\lambda$ represents either $\lambda_{bb}$ or
$\lambda_{bf}$.

If $d\neq 2$, Eqs. (\ref{bfmrg21}) and (\ref{bfmrg23}) will have
two fixed points: (i) the Gaussian fixed point
$(r_{\alpha},\lambda_{bb},\lambda_{bf})=(0,0,0)$ and (ii) the
interacting zero-density fixed point
$(r_{\alpha},\lambda_{bb},\lambda_{bf})=(0,\lambda_*,\lambda_*)$
with $\lambda_*=\epsilon$. On the other hand, for $d=2$, the two
fixed points merge into one -- the Gaussian fixed point. Since
$g_{bb}, g_{bf}>0$, in the dilute limit, the zero-density fixed
point will control the low energy physics in one dimension,
whereas in two dimensions it is the Gaussian fixed point which
will govern the low energy physics. Since the low density implies
strong couplings in one dimension, this nontrivial zero-density
fixed point will describe the strong-coupling physics.

The solution for Eq. (\ref{bfmrg23}) in $d<2$ is of the form
\begin{equation}
 \lambda_{bb}(l)=\frac{\lambda_*}{1+c_{bb}e^{-\epsilon l}} \ , ~~
 \lambda_{bf}(l)=\frac{\lambda_*}{1+c_{bf}e^{-\epsilon l}} \ ,
 \label{bfmrg4}
\end{equation}
with
\begin{eqnarray*}
 c_{bb}=\frac{\lambda_*}{K_dm_bg_{bb}a_0^{\epsilon}}-1 \ , ~~
 c_{bf}=\frac{\lambda_*}{K_dm_{bf}g_{bf}a_0^{\epsilon}}-1 \ ,
\end{eqnarray*}
while in $d=2$ it takes the form
\begin{equation}
 \lambda_{bb}(l)=\frac{K_2m_bg_{bb}}{1+K_2m_bg_{bb}l} \ , ~~
 \lambda_{bf}(l)=\frac{K_2m_{bf}g_{bf}}{1+K_2m_{bf}g_{bf}l} \ .
 \label{bfmrg41}
\end{equation}
Scaling stops at $l=l_*$ where Eq. (\ref{bfmrg23}) can not be
applied any more. This occurs when
$r_b(l_*)+r_f(l_*)=1$,\cite{foot2} yielding
\begin{equation}
 e^{l_*}=\frac{1}{\sqrt{m(\mu_b+\mu_f)a_0^2}} \ . \label{bfmrg42}
\end{equation}
From Eq. (\ref{bfmrg42}), we see that the condition
(\ref{dilute1}) implies that $e^{l_*}\gg 1$.

The uniform mixture corresponds to the state with nonvanishing
values of $\rho_b=\langle |\tilde{\phi}(l_*)|^2\rangle$ and
$\rho_f=\langle |\tilde{\psi}(l_*)|^2\rangle$. Within the
framework of the RG, the phase boundary of the uniform mixture is
determined as follows: First of all, we calculate the effective
potential of the order parameter $\rho_b$ at given $\mu_{\alpha}$:
\begin{equation}
 f=\frac{e^{-(d+2)l_*}}{ma_0^{d+2}}\tilde{f}(l_*) \ ,
 \label{bfmfree1}
\end{equation}
where $\tilde{f}(l_*)$ is the effective potential of the renormalized
Hamiltonian at $l=l_*$. Next, as a function of $\rho_b$,
the equilibrium state is determined by the absolute minimum of $f$ at
fixed $\mu_{\alpha}$. Since the prefactor in Eq. (\ref{bfmfree1}) is
always positive and does not depend on $\rho_b$, the absolute minimum
of $f$ corresponds to the absolute minimum of $\tilde{f}(l_*)$.
Therefore, to search for the absolute minimum of $f$, it suffices to
consider $\tilde{f}(l_*)$. In particular, the stability of the uniform
mixture can be determined from $\tilde{f}(l_*)$. Because the coupling
constants at $l=l_*$ are small (in the sense of the $\epsilon$ expansion
for $d<2$), we may calculate $\tilde{f}(l_*)$ in terms of the perturbation
theory in $\lambda (l_*)$. The leading order result will given by the
mean-field theory. In the following, we will employ this RG-improved
mean-field theory to study the stability of the uniform mixture.
To avoid the confusion, we emphasize again that $f$ is a functional of
the order parameter $\rho_b$ at given $\mu_{\alpha}$, and its absolute
minimum determines the thermodynamical relation between $\rho_b$ and
$\mu_{\alpha}$ and the associated stability conditions. The free energy
density $\Omega$ is obtained by inserting the resulting value of $\rho_b$
into $f$, i.e. $\Omega=f(\mu_{\alpha},\rho_b(\mu_{\alpha}))$.

\subsection{The uniform mixture in one dimension}

We first study the stability of the $1$D uniform mixture. From
Eqs. (\ref{m2}) and (\ref{m3}), to the leading order, $\rho_b$ and
$\rho_f$ are determined by the mean-field equations
\begin{eqnarray*}
 \frac{\lambda^*_{bf}}{K_d\tilde{m}_{bf}}\rho_f=r^*_b
 -\frac{\lambda^*_{bb}}{K_d\tilde{m}_b}\rho_b \ ,
\end{eqnarray*}
and
\begin{eqnarray*}
 \rho_f=\frac{K_d}{d} \! \left(2\tilde{m}_fr^*_f-\frac{(m_b+m_f)\lambda^*_{bf}}
 {K_dm_b}\rho_b\right)^{d/2} ,
\end{eqnarray*}
where $r^*_{\alpha}=r_{\alpha}(l_*)$, $\lambda^*_{bb}=\lambda_{bb}(l_*)$,
and $\lambda^*_{bf}=\lambda_{bf}(l_*)$. In the dilute limit, we may
simply set $\lambda_{bb}^*=\lambda_*=\lambda_{bf}^*$. Hence, by solving
the two equations within the $\epsilon$ expansion, we get
\begin{eqnarray}
 \rho_b &=& \frac{K_d\tilde{m}_b}{\lambda_*}r_b^*-\frac{K_d}{d}(\tilde{m}_b
 +\tilde{m}_f)r_f^*+O(\epsilon) \ , \nonumber \\
 \rho_f &=& \frac{2K_d\tilde{m}_f}{d}r_f^*-\frac{K_d}{d}(\tilde{m}_b
 +\tilde{m}_f)r_b^*+O(\epsilon) \ . \label{1dbfmn1}
\end{eqnarray}

We still have to relate $\mu_{\alpha}$ (or $r_b^*$ and $r_f^*$) to the boson
density $n_b$ and the fermion density $n_f$. This can be achieved through the
scaling equations
\begin{eqnarray}
 & & n_b=e^{-dl_*}n_b(l_*)=e^{-dl_*}a_0^{-d}\rho_b \ , \nonumber \\
 & & n_f=e^{-dl_*}n_f(l_*)=e^{-dl_*}a_0^{-d}\rho_f \ .
     \label{1dbfmn4}
\end{eqnarray}
Inserting Eqs. (\ref{bfmrg42}) and (\ref{1dbfmn1}) into Eq.
(\ref{1dbfmn4}) and noticing that
$r_{\alpha}^*=\mu_{\alpha}/(\mu_b+\mu_f)$, we find that
\begin{equation}
 e^{l_*}= \! \left[\frac{\lambda_*}{2K_d\tilde{m}_f}(1+3\xi)n_ba_0^d+\frac{d(d+2)}
 {4K_d\tilde{m}_f}n_fa_0^d\right]^{-\frac{1}{d}} , \label{bfmrg43}
\end{equation}
and
\begin{eqnarray}
 r_b^* &=& \frac{2\xi n_b+(1+\xi)n_f}{(1+3\xi)n_b+[d(d+2)/(2\lambda_*)]n_f}
 \ , \nonumber \\
 r_f^* &=& \frac{(1+\xi)n_b+[d(d+2)/(4\lambda_*)]n_f}
 {(1+3\xi)n_b+[d(d+2)/(2\lambda_*)]n_f} \ , \label{1dbfmn3}
\end{eqnarray}
where $\xi =m_f/m_b$. By comparing Eq. (\ref{bfmrg42}) with Eq.
(\ref{bfmrg43}), we see that the condition (\ref{dilute1}) indeed
amounts to the diluteness condition $n_{\alpha}a_0^d\ll 1$.

Now we are able to study the stability condition for the $1$D uniform
mixture. From Eq. (\ref{bfmst1}), the stability condition can be written
as
\begin{equation}
 \rho_f^{\epsilon/d}> \! \left(\frac{K_d}{d}\right)^{\epsilon/d} \!
 \frac{m_bm_f(\lambda^*_{bf})^2}{m_{bf}^2\lambda^*_{bb}} \ .
 \label{1dbfmst2}
\end{equation}
Inserting Eqs. (\ref{bfmrg4}), (\ref{1dbfmn4}), and
(\ref{bfmrg43}) into Eq. (\ref{1dbfmst2}), the stability condition
can be written as
\begin{equation}
 1>\lambda_*G(n_b/n_f) \! \left[\frac{1+c_{bb}e^{-\epsilon l_*}}
 {(1+c_{bf}e^{-\epsilon l_*})^2}\right] , \label{1dbfmst3}
\end{equation}
where
\begin{eqnarray*}
 G(x)=\frac{(1+\xi)^2}{4\xi} \! \left[\frac{d+2}{2}
 +\frac{\lambda_*}{d}(1+3\xi)x\right]^{\frac{\epsilon}{d}} .
\end{eqnarray*}
We may use Eq. (\ref{1dbfmst3}) to determine the critical strength
$g_{bf}^{(c)}(\epsilon)$ of $g_{bf}$ at given $g_{bb}$, $n_b$, and
$n_f$ by assuming that $\epsilon\ll 1$. Then, the critical strength
of $g_{bf}$ in $d=1$ can be estimated by setting $\epsilon =1$. To
proceed, we notice that $e^{-\epsilon l_*}$ must be treated as an
exponentially small quantity which can not be expanded as a power
series in $\epsilon$. This is because Eq. (\ref{1dbfmst3}) are
obtained by the double expansion in $e^{-\epsilon l_*}$ and
$\epsilon$.\cite{foot3} Moreover, $G(n_b/n_f),c_{bb}=O(1)\ll
1/\epsilon$ for given $g_{bb}$, $n_b$, and $n_f$. Keeping these in
mind, we find that
\begin{equation}
 K_dm_{bf}g_{bf}a_0^{\epsilon}> -\epsilon e^{-\epsilon l_*} \!
 \left[1+\sqrt{\epsilon G(n_b/n_f)}\right] . \label{1dbfmst31}
\end{equation}
Since the value of the R.H.S. in the inequality (\ref{1dbfmst31})
is negative, it is always satisfied as long as $g_{bf}>0$. This
means that the uniform mixture in $2-\epsilon$ dimensions is
always stable for any finite strength of $g_{bf}>0$. By
extrapolating this result to $\epsilon=1$, our analysis suggests
that the $1$D uniform mixture may be stable against de-mixing for
any finite strength of repulsions.

The action $S$ [Eq. (\ref{bfms1})] can be exactly solved by the
Bethe ansatz under the conditions $m_b=m_f$ and
$g_{bb}=g_{bf}$\cite{bethe}. In the solvable limit, the uniform BF
mixture is shown to be stable against de-mixing. This result is
correctly reproduced in our approach, and is extended to the
region beyond the solvable limit.

\subsection{The uniform mixture in two dimensions}

Now we turn to the $2$D case. Here we will study two problems: (i)
the stability condition for the uniform mixture and (ii) the effect
of fermions on the critical temperature $T_c$ of the superfluid (SF)
phase.

We first examine the stability condition. Following from Eq.
(\ref{1dbfmst2}), the stability condition in $d=2$ becomes
\begin{equation}
 (\lambda_{bf}^*)^2<\frac{2m_{bf}}{m_b+m_f}\lambda_{bb}^* \ .
 \label{2dbfmst1}
\end{equation}
The diluteness condition only requires that $e^{l_*}\gg 1$. On
account of the logarithmic RG flow for the coupling constants in
$d=2$, we have two situations:

When $e^{l_*}\gg 1$ but $l_*=O(1)$, we may set the coupling
constants by their bare values in Eq. (\ref{2dbfmst1}), i.e.
$\lambda_{bb}^*=K_2m_bg_{bb}$ and $\lambda_{bf}^*=K_2m_{bf}g_{bf}$.
Thus, the stability condition in this case is
\begin{equation}
 g_{bf}^2<\frac{2\pi g_{bb}}{m_f} \ . \label{2dbfmst2}
\end{equation}
This is just the mean-field result.

On the other hand, when $l_*\gg 1$, which is referred to as the
extremely dilute limit, we have to insert Eq. (\ref{bfmrg41}) into
Eq. (\ref{2dbfmst1}), yielding
\begin{equation}
 l_*>\frac{(m_b+m_f)^2}{4m_bm_f} \! \left[1-\frac{8\pi}{(m_b+m_f)g_{bf}}
 \right] . \label{2dbfmst3}
\end{equation}
We still have to determine $l_*$ as a function of $n_{\alpha}$. For
$l_*\gg 1$, we may simply set $\lambda_{bb}^*=1/l_*=\lambda_{bf}^*$.
The rest of the procedure is similar to the $1$D case, and the
result is
\begin{eqnarray*}
 l_*\approx\frac{1}{2}\ln{ \! \! \left(\frac{1+3\xi}{2\pi n_fa_0^d}\right)}
 ,
\end{eqnarray*}
to the leading order in the $1/l_*$ expansion. Consequently, the
extremely dilute limit is given by the condition
\begin{equation}
 \ln{ \! \! \left(\frac{1}{2\pi n_fa_0^d}\right)}\gg 1 \ .
 \label{2dbfmedl}
\end{equation}
Since the right hand side of Eq. (\ref{2dbfmst3}) is of $O(1)$, it
is automatically satisfied as long as Eq. (\ref{2dbfmedl}) is
obeyed. Therefore, we conclude that in the extremely dilute limit
the uniform mixture is stable against demixing for any finite strength
of repulsions. In other words, the quantum fluctuations stabilize the
uniform mixture with repulsive interactions in the extremely dilute
limit.

In the parameter space where the uniform mixture is stable, the
system will becomes a superfluid (SF) phase through the KT-type
transition as lowering the temperature $T$. Here we would like to
calculate the critical temperature $T_c$ for the SF phase in terms
of the RG. Now there are three relevant parameters $r_b(l)$,
$r_f(l)$, and $t_l$ in the sense of RG, where the dimensionless
temperature $t_l$ is defined as $t_l=t_0e^{2l}$ with
$t_0=mTa_0^2$. To proceed, we will assume that $|\mu_{\alpha}|\ll
T_c\ll 1/(m_{\alpha}a_0^2)$. In this parameter regime, it is $t_l$
which reaches $O(1)$ first under the RG transformations. Hence, we
run the RG equations to the scale $l=\tilde{l}$ such that
$t_{\tilde{l}}=1$. Accordingly, $\tilde{l}$ is given by
\begin{equation}
 \tilde{l}=\frac{1}{2}\ln{\! \! \left[\frac{1}{mTa_0^2}\right]} .
 \label{bfmtrg}
\end{equation}
Before $l$ reaches $\tilde{l}$, $t_l\ll 1$, and thus we may
approximate the RG equations for $r_{\alpha}(l)$, $\lambda_{bb}(l)$,
and $\lambda_{bf}(l)$ as the ones at $T=0$. This may introduce
multiplicative errors of order unity coming from the imprecise
treatment of the regime $t_l\sim 1$, but makes the analysis simpler.
At $l=\tilde{l}$, the action $S$ becomes
\begin{eqnarray*}
 S &=& \! \int^1_0 \! \! d\tilde{\tau} \! \! \int \! \! d^2\tilde{x}
 ~\tilde{\psi}^{\dagger} \! \! \left[\partial_{\tilde{\tau}}
 -\frac{\nabla^2}{2\tilde{m}_f}-\tilde{r}_f\right] \! \tilde{\psi}
 \\
 & & + \! \int^1_0 \! \! d\tilde{\tau} \! \! \int \! \! d^2\tilde{x}
 ~\tilde{\phi}^{\dagger} \! \! \left[\partial_{\tilde{\tau}}
 -\frac{\nabla^2}{2\tilde{m}_b}-\tilde{r}_b\right] \! \tilde{\phi}
 \\
 & & + \! \int^1_0 \! \! d\tilde{\tau} \! \! \int \! \! d^2\tilde{x}
 \! \left[\frac{\tilde{\lambda}}{2K_2\tilde{m}_b}|\tilde{\phi}|^4
 +\frac{\tilde{\lambda}}{K_2\tilde{m}_{bf}}|\tilde{\phi}|^2
 |\tilde{\psi}|^2\right] ,
\end{eqnarray*}
where $\tilde{r}_{\alpha}=r_{\alpha}(\tilde{l})=\mu_{\alpha}/T$ and
$\tilde{\lambda}=1/\tilde{l}$. Here we assume that $\tilde{l}\gg 1$
so that $\lambda_{bb}(\tilde{l})\approx \tilde{\lambda}$ and
$\lambda_{bf}(\tilde{l})\approx \tilde{\lambda}$. Since
$\tilde{\lambda}\ll 1$, a perturbative expansion in
$\tilde{\lambda}$ is reliable.

The field $\tilde{\phi}$ may be considered the order parameter of
the SF phase. Hence, the critical temperature can be determined by
the equation
\begin{equation}
 \tilde{r}_b+\tilde{\Sigma}=0 \ , \label{2dbfmtc}
\end{equation}
where $\tilde{\Sigma}$ is the self-energy of $\tilde{\phi}$. To the
one-loop order, $\tilde{\Sigma}$ is given by
\begin{eqnarray*}
 \tilde{\Sigma}\approx 2\tilde{\lambda}\ln{\tilde{r}_b} \ .
\end{eqnarray*}
Here, we keep only the leading term in the small
$\tilde{r}_{\alpha}$ expansion. Inserting this result into Eq.
(\ref{2dbfmtc}), we obtain
\begin{equation}
 \tilde{r}_b\approx -2\tilde{\lambda}\ln{(2\tilde{\lambda})} \ .
 \label{2dbfmtc2}
\end{equation}

We have to express $\tilde{r}_b$ as a function of $n_{\alpha}$ and
$T$. The procedure is similar to the $1$D case, and the result is
\begin{eqnarray*}
 \tilde{r}_b=\frac{\tilde{\lambda}}{K_2m_bT}n_b+\frac{\tilde{\lambda}}
 {K_2m_{bf}T}n_f \ .
\end{eqnarray*}
Inserting this expression and using Eq. (\ref{bfmtrg}), we find
\begin{equation}
 \frac{\pi}{m_bT} \! \left(n_b+\frac{1+\xi}{2\xi}n_f\right) \!
 =\ln{ \! \left\{\frac{1}{4}\ln{ \! \! \left[\frac{1}{mTa_0^2}\right]}\right\}}
 . \label{2dbfmtc3}
\end{equation}
From Eq. (\ref{2dbfmtc3}), $T_c$ is given by
\begin{equation}
 T_c \! = \! \frac{\pi [n_b+(1+\xi)/(2\xi)n_f]}
 {m_b \! \ln{ \! \! \left\{ \! \frac{1}{4} \! \ln{ \! \! \left[\frac{\tilde{m}_b}
 {\pi [n_b+(1+\xi)/(2\xi)n_f]a_0^2}\right]} \! \right\}}}
 \ . \label{2dbfmtc31}
\end{equation}
By setting $n_f=0$, Eq. (\ref{2dbfmtc31}) reduces to the critical
temperature for a single-component dilute Bose gas.\cite{FH} Equation
(\ref{2dbfmtc31}) suggests that the inclusion of fermions increases the
critical temperature.

\section{The Bose-Bose mixture}
\label{2bm}

The dilute BB mixture can be described by the action
\begin{equation}
 S= \! \int^{\beta}_0 \! \! d\tau \! \! \int \! \! d^dx \! \left[ \!
 \sum_{\alpha =1,2} \! \Psi_{\alpha}^{\dagger} \! \! \left(\partial_{\tau}
 -\frac{\nabla^2}{2m_{\alpha}}-\mu_{\alpha}\right) \! \Psi_{\alpha}+U\right]
 , \label{2boses1}
\end{equation}
where
\begin{eqnarray*}
 U=\frac{1}{2} \! \sum_{\alpha =1,2}u_{\alpha\alpha}|\Psi_{\alpha}|^4
 +u_{12}|\Psi_1|^2|\Psi_2|^2 \ .
\end{eqnarray*}
Here $\Psi_{\alpha}$ are the fields describing the bosonic atoms
with masses $m_{\alpha}$. The couplings $u_{\alpha\beta}$ are
related to the scattering lengths $a_{\alpha\beta}$ through the
relations $u_{12}=2\omega_{\perp}a_{12}$ and
$u_{\alpha\alpha}=2\omega_{\perp}a_{\alpha\alpha}$ for the $1$D
trap, and $u_{12}=\sqrt{4\pi(m_1+m_2)\omega_{\perp}/m_1m_2}a_{12}$
and
$u_{\alpha\alpha}=\sqrt{8\pi\omega_{\perp}/m_{\alpha}}a_{\alpha\alpha}$
for a $2$D trap, where $\omega_{\perp}$ is the transverse
confining frequency. We shall take $u_{\alpha\beta}>0$.

By integrating out the fast modes, the one-loop RG equations in
the dilute limit are completely identical to Eqs. (\ref{bfmrg21})
and (\ref{bfmrg23}), with the initial conditions
\begin{eqnarray*}
 r_{\alpha}(0)=(m_1+m_2)\mu_{\alpha}a_0^2 \ , ~~
 \lambda_{\alpha\beta}(0)=K_dm_{\alpha\beta}u_{\alpha\beta}a_0^{\epsilon}
 \ .
\end{eqnarray*}
Here $m_{\alpha\beta}=2m_{\alpha}m_{\beta}/(m_1+m_2)$. Hence, the
fixed-point structure is identical to the one for the BF mixture. The
solution of the RG equation for
$\lambda_{\alpha\beta}(l)$ in $d=1$ is of the form
\begin{equation}
 \lambda_{\alpha\beta}(l)=\frac{\lambda_*}{1+c_{\alpha\beta}e^{-l}}
 \ , \label{2brg31}
\end{equation}
where
\begin{eqnarray*}
 c_{\alpha\beta}=\frac{\lambda_*}{K_1m_{\alpha\beta}a_0u_{\alpha\beta}}-1
 \ ,
\end{eqnarray*}
while in $d=2$ it is given by
\begin{equation}
 \lambda_{\alpha\alpha}(l)=\frac{K_2m_{\alpha\beta}u_{\alpha\beta}}
 {1+K_2m_{\alpha\beta}u_{\alpha\beta}l} \ .
 \label{2brg32}
\end{equation}
Scaling stops when $l=l_*$, where $r_1(l_*)+r_2(l_*)=1$, yielding
\begin{equation}
 e^{l_*}=\frac{1}{\sqrt{(m_1+m_2)(\mu_1+\mu_2)a_0^2}} \ . \label{2brg4}
\end{equation}
By a procedure similar to the Bose-Fermi mixture, $l_*$ in $d=1$
can be expressed as
\begin{equation}
 e^{l_*}=\frac{m_{12}/(\pi \lambda_*)}{a_0[(m_1+3m_2)n_1+(3m_1+m_2)n_2]} \ ,
 \label{2brg41}
\end{equation}
in the dilute limit. On the other hand, in $d=2$, $l_*$ takes the
form
\begin{equation}
 l_*=\ln{\sqrt{\frac{m_{12}/(2\pi)}{[(m_1+3m_2)n_1+(3m_1+m_2)n_2]a_0^2}}}
 \ , \label{2brg42}
\end{equation}
in the extremely dilute limit. Here the extremely dilute limit is
defined by $l_*\gg 1$. (The diluteness condition only guarantees
that $e^{l_*}\gg 1$.)

We first use the above results to study the $1$D uniform mixture.
Following the argument similar to the dilute BF mixture, the stability
condition for the uniform mixture can be obtained from a mean-field
analysis on the renormalized Hamiltonian, yielding
\begin{equation}
 \frac{4m_1m_2}{(m_1+m_2)^2}\lambda_{11}^*\lambda_{22}^*>(\lambda_{12}^*)^2
 \ , \label{2bst1}
\end{equation}
where $\lambda_{\alpha\beta}^*=\lambda_{\alpha\beta}(l_*)$. It is
clear that the fixed-point Hamiltonian (with
$\lambda_{\alpha\beta}=\lambda_*$) cannot satisfy Eq.
(\ref{2bst1}). This means that the fixed-point Hamiltonian does
not describe a uniform mixture, and thus we cannot take the
continuum limit directly, as we have done for the BF mixture.

To proceed, we insert Eq. (\ref{2brg31}) into Eq. (\ref{2bst1}) and
use Eq. (\ref{2brg41}). Then, in the dilute limit ($e^{l_*}\gg 1$),
the stability condition can be expressed as
\begin{equation}
 0<u_{12}<{\mathcal U}^{(1)}_c \ , \label{2b1d11}
\end{equation}
where
\begin{eqnarray*}
 {\mathcal U}^{(1)}_c=\frac{u_c}{1+\frac{m_1m_2(m_1-m_2)^2u_c}{2\pi^2(m_1+m_2)^2
 [(m_1+3m_2)n_1+(3m_1+m_2)n_2]}} \ ,
\end{eqnarray*}
is the critical strength of $u_{12}$ in $d=1$, and
\begin{eqnarray*}
 u_c=\frac{2m_{12}u_{11}u_{22}}{m_1u_{11}+m_2u_{22}} \ .
\end{eqnarray*}
In the above, we have set $\lambda_*=\epsilon =1$. From the expression
for ${\mathcal U}_c^{(1)}$, we see that for the mixtures consisting of
two different kinds of bosons, i.e. $m_1\neq m_2$, the critical strength
${\mathcal U}_c^{(1)}$ is a linear function of boson densities
\begin{eqnarray*}
 {\mathcal U}^{(1)}_c\approx\frac{2\pi^2(m_1+m_2)^2}{m_1m_2(m_1-m_2)^2}
 [(m_1+3m_2)n_1+(3m_1+m_2)n_2] \ ,
\end{eqnarray*}
in the dilute limit, which may be much smaller than $u_c$. Moreover,
${\mathcal U}^{(1)}_c$ is is insensitive to the values of $u_{\alpha\alpha}$.
On the other hand, for the mixtures composed of two different hyperfine
states of the same boson, i.e. $m_1=m_2$, ${\mathcal U}^{(1)}_c$ reduces to
\begin{eqnarray*}
 {\mathcal U}^{(1)}_c=\frac{2u_{11}u_{22}}{u_{11}+u_{22}} \ ,
\end{eqnarray*}
in the dilute limit, which is independent of the boson densities.

The stability conditions we obtained are valid only in the strong-coupling
regime. We notice that they are different from the result obtained from
bosonization, which gave ${\mathcal U}_c^{(1)}=\pi^2\sqrt{n_1n_1/(m_1m_2)}$.
The study of the $1$D BB mixture with the exchange symmetry, i.e.
$m_1=m_2$ and $u_{11}=u_{22}=U$, by the finite-size density-matrix
RG (FSDMRG) indicates that the phase separation occurs for all
boson densities when $u_{12}>U$.\cite{MPD} Our conclusion is
consistent with this result when the system possesses the exchange
symmetry.

Next, we turn into the $2$D case. Due to the logarithmic RG flow of
the coupling constants, we have two situations: (i) When $e^{l_*}\gg
1$ but $l_*=O(1)$, we may take $\lambda_{\alpha\beta}^*$ by their
initial values, i.e. $\lambda_{\alpha\beta}^*\approx
K_2m_{\alpha\beta}u_{\alpha\beta}$. Inserting these into Eq.
(\ref{2bst1}) gives rise to the mean-field result
\begin{equation}
 0<u_{12}<\sqrt{u_{11}u_{22}} \ . \label{2bst2}
\end{equation}
(ii) In the extremely dilute limit, i.e. $l_*\gg 1$, we insert Eq.
(\ref{2brg32}) into Eq. (\ref{2bst1}). Then, the stability
condition in this situation becomes
\begin{equation}
 0<u_{12}<{\mathcal U}_c^{(2)} \ , \label{2b2dst2}
\end{equation}
where
\begin{eqnarray*}
 {\mathcal U}_c^{(2)}=\frac{u_c}{1+\frac{(m_1-m_2)^2u_c}
 {16\pi(m_1+m_2)}l_*} \ ,
\end{eqnarray*}
and $l_*$ is given by Eq. (\ref{2brg42}). We see that for the
mixtures composed of two different hyperfine states of the same
boson, i.e. $m_1=m_2$, the critical strength is given by
${\mathcal U}_c^{(2)}=u_c$, which is independent of the boson
densities, similar to the $1$D case. On the other hands, for the
mixtures consisting of two different kinds of bosons, i.e.
$m_1\neq m_2$, ${\mathcal U}_c^{(2)}\approx 16\pi
(m_1+m_2)/[(m_1-m_2)^2l_*]$. That is, it is a function of boson
densities, which is insensitive to the values of
$u_{\alpha\alpha}$. However, the dependence of ${\mathcal
U}_c^{(2)}$ on the boson densities is much weaker than the $1$D
case due to the presence of the logarithm. This is a
characteristic of the $2$D Bose gas.

A schematic phase diagram for the $1$D and $2$D BB mixtures with
repulsions is shown in Fig. \ref{1dbbp}, where we have taken
$n_1=n/2=n_2$. For $m_1\neq m_2$, ${\mathcal U}_c^{(1)}$ is a
linear function of $n$ at low density and ${\mathcal U}_c^{(2)}$
is a linear function of $1/\ln{[1/(\eta na_0^2)]}$ at extremely
low density with $\eta=\frac{m_2/m_1}{2\pi (1+m_2/m_1)^2}$. On the
other hand, ${\mathcal U}^{(1)}_c=u_c={\mathcal U}^{(2)}_c$ in the
limit $na_0^d\rightarrow 0$ for $m_1=m_2$. In both cases,
${\mathcal U}_c^{(d)}$ approaches the mean-field value
$u_{mf}=\sqrt{u_{11}u_{22}}$ at high density. We notice that the
action $S$ [Eq. (\ref{2boses1})] possesses an SU($2$) symmetry
when $m_1=m_2$, $\mu_1=\mu_2$, and $u_{11}=u_{22}=u_{12}$. In this
SU($2$) symmetrical limit, $u_c=u_{mf}$, and the phase boundary
may coincide with the mean-field prediction.

\section{Conclusions and discussions}

We employ the RG to study the dilute BF and BB mixtures with
repulsive interactions in low dimensions. For the $1$D BF mixture,
we show that the uniform mixture is stable against de-mixing for
any finite strength of repulsions. This conclusion is consistent
with the prediction obtained from the Bethe ansatz when the system
can be solved exactly. The bosonization approach gave a stability
condition of the mean-field type at large boson densities (the
weak-coupling regime)\cite{CH}, while at low boson densities (the
strong-coupling regime) the stability condition becomes
$0<u_{12}<\pi^2\sqrt{n_bn_f/(m_bm_f)}$. This result is inconsistent
with the exact solution, while in the solvable limit ours is
consistent with the Bethe-ansatz result. We believe that this
inconsistency arises from the fact that the relations between the
Luttinger liquid parameters and the short-distance parameters like
$g_{bb}$ and $g_{bf}$ are not reliable in the dilute limit where
the system goes into the strong-coupling regime.

For the $2$D BF mixture with repulsions in the extremely dilute
limit, the uniform mixture is also stable irrespective of the
interaction strength. The mean-field result is expected to be
valid in the moderate and high density regime. We also calculate
the critical temperature $T_c$ for the SF phase. By comparing with
$T_c$ for the single-component dilute Bose gas, we find that the
inclusion of fermions increases $T_c$. That is, the presence of
fermions enhances the long range SF ordering.

For the $1$D BB mixture, the low-energy physics is controlled by a
fixed point which does not support the uniform mixture. This
observation immediately leads to two consequences: (i) First of
all, the equilibrium properties of the uniform mixture will not
exhibit universal behaviors even in the strong-coupling regime
because we cannot take the continuum limit. (ii) Next, the region
of the stable uniform mixture with $m_1\neq m_2$ can be very
narrow in the dilute limit. For the $2$D BB mixture in the
extremely dilute limit, the region of the stable uniform mixture
with $m_1\neq m_2$ is also narrow because the interactions are
strongly renormalized in this limit. The mean-field result can be
applied only in the moderate and high density regime.

A recent paper also studied the stability conditions for the
low-dimensional dilute BB mixture in terms of the RG
approach\cite{K}. The main distinction between our results and the
ones in Ref. 16 lies at the $1$D case with $m_1\neq m_2$, where
our predicted value of ${\mathcal U}_c^{(1)}$ can be much smaller
than $u_c$ in the dilute limit (our $u_c$ is basically $u^{(1d)}$
in Ref. 16), while in one dimension with $m_1=m_2$ and in two
dimensions, both are qualitatively consistent with each other. The
discrepancy may be originated from the different criteria of the
stability: In Ref. 16, the criterion for the stability is the
positive definiteness of the potential term in the action at any
scale $l<l^*$. In our opinion, the use of the positive
definiteness of the potential term in the action amounts to some
kind of mean-field treatment, and such a procedure is meaningful
only when the quantum fluctuations have been properly taken into
account. Hence, this criterion can be used only at the scale
$l=l^*$ where the correlation length becomes of $O(1)$ in the unit
of the short-distance cutoff at $l=l^*$. This is the criterion
which we employed in this work. The issue about whether or not the
critical strength for the $1$D uniform BB mixture can be very
small in the dilute limit should be answered by further numerical
or experimental studies.

\acknowledgments

The work of Y.-W. Lee is supported by the National Science Council
of Taiwan under grant NSC 99-2112-M-029-004-MY3. The work of Y.L.
Lee is supported by the National Science Council of Taiwan under
grant NSC 98-2112-M-018-003-MY3.

\appendix
\section{Mean-field theory of the uniform Bose-Fermi mixture in $d$ dimensions}
\label{ap1}

By taking the ansatz $\rho =|\phi|^2$, The mean-field free energy
density $f$ for the action $S$ [Eq. (\ref{bfms1})] is of the form
\begin{eqnarray*}
 f &=& -\mu_b\rho+\frac{g_{bb}}{2}\rho^2-\frac{2K_d(2m_f)^{d/2}}
 {d(d+2)}\Theta (\mu_f-g_{bf}\rho) \\
 & & \times (\mu_f-g_{bf}\rho)^{d/2+1} \ ,
\end{eqnarray*}
where $\Theta (x)=0,1$ for $x<0$ and $x>0$, respectively. The
value of $\rho$ is determined by the minimum of $f$, which leads
to the mean-field equation
\begin{eqnarray*}
 \frac{K_dg_{bf}}{d}\Theta (\mu_f-g_{bf}\rho)[2m_f(\mu_f-g_{bf}
 \rho)]^{d/2}=\mu_b-g_{bb}\rho \ .
\end{eqnarray*}

The uniform mixture corresponds to the solution of the mean-field
equation with $0<\rho<\mu_f/g_{bf}$. Hence, for the uniform
mixture, the free energy density is of the form
\begin{equation}
 f_M=-\mu_b\rho+\frac{g_{bb}}{2}\rho^2-\frac{2K_d(2m_f)^{d/2}}
 {d(d+2)}(\mu_f-g_{bf}\rho)^{d/2+1} \ , \label{m1}
\end{equation}
where $\rho_b$ is the solution of the equation
\begin{equation}
 \frac{K_dg_{bf}}{d}[2m_f(\mu_f-g_{bf}\rho)]^{d/2}=\mu_b-g_{bb}\rho
 \ , \label{m2}
\end{equation}
satisfying the constraint $0<\rho<\mu_f/g_{bf}$. The density of
fermions $\rho_f$ can be obtained from $f_M$ through the
thermodynamical relation $\rho_f=-\partial f_M/\partial\mu_f$,
yielding
\begin{equation}
 \rho_f=\frac{K_d}{d}[2m_f(\mu_f-g_{bf}\rho)]^{d/2} \ .
 \label{m3}
\end{equation}

Using the thermodynamical relation $\rho_b=-\partial
f_M/\partial\mu_b$ where $\rho_b$ denotes the boson density, one may
find that $\rho=\rho_b$. In terms of this relation and Eqs.
(\ref{m2}), (\ref{m3}) to eliminate $\mu_{\alpha}$, one may express
$f_M$ as a function of $\rho_{\alpha}$. Using Euler's relation
$f_M=-P$ and the thermodynamical relation $P=-\partial E/\partial
V$, one may obtain the total energy density
\begin{equation}
 \frac{E_M}{V}=\frac{g_{bb}}{2}\rho_b^2+g_{bf}\rho_b\rho_f+\frac{d}
 {d+2}g_{ff}\rho_f^{1+2/d} \ , \label{bfme1}
\end{equation}
where
\begin{eqnarray*}
 g_{ff}=\frac{1}{2m_f} \! \left(\frac{d}{K_d}\right)^{2/d} .
\end{eqnarray*}
By considering small density fluctuations, we obtain from Eq.
(\ref{bfme1}) the linear stability condition
\begin{equation}
 g_{bf}^2<\frac{2g_{ff}}{d}g_{bb}n_f^{2/d-1} \ . \label{bfmst1}
\end{equation}



\begin{thebibliography}{99}
\bibitem{foot2} Strictly speaking, we should define $l_*$ by
         $r_b(l_*)+r_f(l_*)=\alpha$ where $\alpha =O(1)$. However, $\alpha$ always
         appears in $e^{l_*}$ in the guise of $a_0/\alpha$. This means that the exact
         value of $\alpha$ depends on the choice of the short-distance cutoff $a_0$,
         which does not affect our results, such as the dependence of the physical
         quantities on $n_b$ and $n_f$.
\bibitem{foot3} The condition $e^{\epsilon l_*}\gg 1$ insures that we are adressing
         the physics near the interacting zero-density fixed point, i.e. the low-energy
         physics in the strong-coupling regime. In particular, we obtain Eqs.
         (\ref{1dbfmst3}) or (\ref{1dbfmst31}) by assuming $e^{-\epsilon l_*}\ll 1$.
         Hence, a naive substitution of $\epsilon=0$ into them will give rise to
         $e^{\epsilon l_*}\rightarrow 1$ and contradict this condition. Therefore,
         these results in $2-\epsilon$ dimensions can not be used in $2$D, which has
         a completely different fixed-point structure.
\end{thebibliography}
\end{document}